\documentclass[aps,pre,epsfigure,twocolumn,superscriptaddress,nofootinbib]{revtex4-1}
\usepackage[utf8]{inputenc} 
\usepackage{amsmath}
\usepackage{mathrsfs}
\usepackage{graphicx,epstopdf}
\usepackage{amssymb, theorem, xspace}
\usepackage{float}
\usepackage{dcolumn}
\usepackage{slashed}
\usepackage{tikz}
\usetikzlibrary{quantikz}
\usepackage[colorlinks=true,linkcolor=blue,urlcolor=blue,citecolor=blue,pdfusetitle]{hyperref}
\usepackage[utf8]{inputenc}
\usepackage[english]{babel}
\usepackage{subfigure}
\usepackage{blindtext}
\usepackage{amsfonts}
\usepackage{bbm}
\usepackage{enumerate}
\usepackage{color}
\usepackage{latexsym}
\usepackage{physics}
\usepackage{times,txfonts}

\newcommand{\Ucal}{\mathcal{U}}

\newcommand{\1}{\mathbbm{1}}

\usepackage[breakable]{tcolorbox}
    \makeatletter
    \newcommand{\boxspacing}{\kern\kvtcb@left@rule\kern\kvtcb@boxsep}
    \makeatother

 \definecolor{incolor}{HTML}{303F9F}
    \definecolor{outcolor}{HTML}{D84315}
    \definecolor{cellborder}{HTML}{CFCFCF}
    \definecolor{cellbackground}{HTML}{F7F7F7}
  
 \usepackage{upquote} 
    \usepackage{eurosym} 
    \usepackage[mathletters]{ucs} 
    \usepackage{fancyvrb} 
    \usepackage{grffile} 
    \makeatletter 
    \def\Gread@@xetex#1{%
      \IfFileExists{"\Gin@base".bb}%
      {\Gread@eps{\Gin@base.bb}}%
      {\Gread@@xetex@aux#1}%
    }
    \makeatother
 
    \DefineVerbatimEnvironment{Highlighting}{Verbatim}{commandchars=\\\{\}}

    \DefineVerbatimEnvironment{Highlighting}{Verbatim}{commandchars=\\\{\}}

    \let\Oldtex\TeX
    \let\Oldlatex\LaTeX
    \renewcommand{\TeX}{\textrm{\Oldtex}}
    \renewcommand{\LaTeX}{\textrm{\Oldlatex}}
    \title{artigo_qiskit}

\makeatletter
\def\PY@reset{\let\PY@it=\relax \let\PY@bf=\relax%
    \let\PY@ul=\relax \let\PY@tc=\relax%
    \let\PY@bc=\relax \let\PY@ff=\relax}
\def\PY@tok#1{\csname PY@tok@#1\endcsname}
\def\PY@toks#1+{\ifx\relax#1\empty\else%
    \PY@tok{#1}\expandafter\PY@toks\fi}
\def\PY@do#1{\PY@bc{\PY@tc{\PY@ul{%
    \PY@it{\PY@bf{\PY@ff{#1}}}}}}}
\def\PY#1#2{\PY@reset\PY@toks#1+\relax+\PY@do{#2}}

\expandafter\def\csname PY@tok@w\endcsname{\def\PY@tc##1{\textcolor[rgb]{0.73,0.73,0.73}{##1}}}
\expandafter\def\csname PY@tok@c\endcsname{\let\PY@it=\textit\def\PY@tc##1{\textcolor[rgb]{0.25,0.50,0.50}{##1}}}
\expandafter\def\csname PY@tok@cp\endcsname{\def\PY@tc##1{\textcolor[rgb]{0.74,0.48,0.00}{##1}}}
\expandafter\def\csname PY@tok@k\endcsname{\let\PY@bf=\textbf\def\PY@tc##1{\textcolor[rgb]{0.00,0.50,0.00}{##1}}}
\expandafter\def\csname PY@tok@kp\endcsname{\def\PY@tc##1{\textcolor[rgb]{0.00,0.50,0.00}{##1}}}
\expandafter\def\csname PY@tok@kt\endcsname{\def\PY@tc##1{\textcolor[rgb]{0.69,0.00,0.25}{##1}}}
\expandafter\def\csname PY@tok@o\endcsname{\def\PY@tc##1{\textcolor[rgb]{0.40,0.40,0.40}{##1}}}
\expandafter\def\csname PY@tok@ow\endcsname{\let\PY@bf=\textbf\def\PY@tc##1{\textcolor[rgb]{0.67,0.13,1.00}{##1}}}
\expandafter\def\csname PY@tok@nb\endcsname{\def\PY@tc##1{\textcolor[rgb]{0.00,0.50,0.00}{##1}}}
\expandafter\def\csname PY@tok@nf\endcsname{\def\PY@tc##1{\textcolor[rgb]{0.00,0.00,1.00}{##1}}}
\expandafter\def\csname PY@tok@nc\endcsname{\let\PY@bf=\textbf\def\PY@tc##1{\textcolor[rgb]{0.00,0.00,1.00}{##1}}}
\expandafter\def\csname PY@tok@nn\endcsname{\let\PY@bf=\textbf\def\PY@tc##1{\textcolor[rgb]{0.00,0.00,1.00}{##1}}}
\expandafter\def\csname PY@tok@ne\endcsname{\let\PY@bf=\textbf\def\PY@tc##1{\textcolor[rgb]{0.82,0.25,0.23}{##1}}}
\expandafter\def\csname PY@tok@nv\endcsname{\def\PY@tc##1{\textcolor[rgb]{0.10,0.09,0.49}{##1}}}
\expandafter\def\csname PY@tok@no\endcsname{\def\PY@tc##1{\textcolor[rgb]{0.53,0.00,0.00}{##1}}}
\expandafter\def\csname PY@tok@nl\endcsname{\def\PY@tc##1{\textcolor[rgb]{0.63,0.63,0.00}{##1}}}
\expandafter\def\csname PY@tok@ni\endcsname{\let\PY@bf=\textbf\def\PY@tc##1{\textcolor[rgb]{0.60,0.60,0.60}{##1}}}
\expandafter\def\csname PY@tok@na\endcsname{\def\PY@tc##1{\textcolor[rgb]{0.49,0.56,0.16}{##1}}}
\expandafter\def\csname PY@tok@nt\endcsname{\let\PY@bf=\textbf\def\PY@tc##1{\textcolor[rgb]{0.00,0.50,0.00}{##1}}}
\expandafter\def\csname PY@tok@nd\endcsname{\def\PY@tc##1{\textcolor[rgb]{0.67,0.13,1.00}{##1}}}
\expandafter\def\csname PY@tok@s\endcsname{\def\PY@tc##1{\textcolor[rgb]{0.73,0.13,0.13}{##1}}}
\expandafter\def\csname PY@tok@sd\endcsname{\let\PY@it=\textit\def\PY@tc##1{\textcolor[rgb]{0.73,0.13,0.13}{##1}}}
\expandafter\def\csname PY@tok@si\endcsname{\let\PY@bf=\textbf\def\PY@tc##1{\textcolor[rgb]{0.73,0.40,0.53}{##1}}}
\expandafter\def\csname PY@tok@se\endcsname{\let\PY@bf=\textbf\def\PY@tc##1{\textcolor[rgb]{0.73,0.40,0.13}{##1}}}
\expandafter\def\csname PY@tok@sr\endcsname{\def\PY@tc##1{\textcolor[rgb]{0.73,0.40,0.53}{##1}}}
\expandafter\def\csname PY@tok@ss\endcsname{\def\PY@tc##1{\textcolor[rgb]{0.10,0.09,0.49}{##1}}}
\expandafter\def\csname PY@tok@sx\endcsname{\def\PY@tc##1{\textcolor[rgb]{0.00,0.50,0.00}{##1}}}
\expandafter\def\csname PY@tok@m\endcsname{\def\PY@tc##1{\textcolor[rgb]{0.40,0.40,0.40}{##1}}}
\expandafter\def\csname PY@tok@gh\endcsname{\let\PY@bf=\textbf\def\PY@tc##1{\textcolor[rgb]{0.00,0.00,0.50}{##1}}}
\expandafter\def\csname PY@tok@gu\endcsname{\let\PY@bf=\textbf\def\PY@tc##1{\textcolor[rgb]{0.50,0.00,0.50}{##1}}}
\expandafter\def\csname PY@tok@gd\endcsname{\def\PY@tc##1{\textcolor[rgb]{0.63,0.00,0.00}{##1}}}
\expandafter\def\csname PY@tok@gi\endcsname{\def\PY@tc##1{\textcolor[rgb]{0.00,0.63,0.00}{##1}}}
\expandafter\def\csname PY@tok@gr\endcsname{\def\PY@tc##1{\textcolor[rgb]{1.00,0.00,0.00}{##1}}}
\expandafter\def\csname PY@tok@ge\endcsname{\let\PY@it=\textit}
\expandafter\def\csname PY@tok@gs\endcsname{\let\PY@bf=\textbf}
\expandafter\def\csname PY@tok@gp\endcsname{\let\PY@bf=\textbf\def\PY@tc##1{\textcolor[rgb]{0.00,0.00,0.50}{##1}}}
\expandafter\def\csname PY@tok@go\endcsname{\def\PY@tc##1{\textcolor[rgb]{0.53,0.53,0.53}{##1}}}
\expandafter\def\csname PY@tok@gt\endcsname{\def\PY@tc##1{\textcolor[rgb]{0.00,0.27,0.87}{##1}}}
\expandafter\def\csname PY@tok@err\endcsname{\def\PY@bc##1{\setlength{\fboxsep}{0pt}\fcolorbox[rgb]{1.00,0.00,0.00}{1,1,1}{\strut ##1}}}
\expandafter\def\csname PY@tok@kc\endcsname{\let\PY@bf=\textbf\def\PY@tc##1{\textcolor[rgb]{0.00,0.50,0.00}{##1}}}
\expandafter\def\csname PY@tok@kd\endcsname{\let\PY@bf=\textbf\def\PY@tc##1{\textcolor[rgb]{0.00,0.50,0.00}{##1}}}
\expandafter\def\csname PY@tok@kn\endcsname{\let\PY@bf=\textbf\def\PY@tc##1{\textcolor[rgb]{0.00,0.50,0.00}{##1}}}
\expandafter\def\csname PY@tok@kr\endcsname{\let\PY@bf=\textbf\def\PY@tc##1{\textcolor[rgb]{0.00,0.50,0.00}{##1}}}
\expandafter\def\csname PY@tok@bp\endcsname{\def\PY@tc##1{\textcolor[rgb]{0.00,0.50,0.00}{##1}}}
\expandafter\def\csname PY@tok@fm\endcsname{\def\PY@tc##1{\textcolor[rgb]{0.00,0.00,1.00}{##1}}}
\expandafter\def\csname PY@tok@vc\endcsname{\def\PY@tc##1{\textcolor[rgb]{0.10,0.09,0.49}{##1}}}
\expandafter\def\csname PY@tok@vg\endcsname{\def\PY@tc##1{\textcolor[rgb]{0.10,0.09,0.49}{##1}}}
\expandafter\def\csname PY@tok@vi\endcsname{\def\PY@tc##1{\textcolor[rgb]{0.10,0.09,0.49}{##1}}}
\expandafter\def\csname PY@tok@vm\endcsname{\def\PY@tc##1{\textcolor[rgb]{0.10,0.09,0.49}{##1}}}
\expandafter\def\csname PY@tok@sa\endcsname{\def\PY@tc##1{\textcolor[rgb]{0.73,0.13,0.13}{##1}}}
\expandafter\def\csname PY@tok@sb\endcsname{\def\PY@tc##1{\textcolor[rgb]{0.73,0.13,0.13}{##1}}}
\expandafter\def\csname PY@tok@sc\endcsname{\def\PY@tc##1{\textcolor[rgb]{0.73,0.13,0.13}{##1}}}
\expandafter\def\csname PY@tok@dl\endcsname{\def\PY@tc##1{\textcolor[rgb]{0.73,0.13,0.13}{##1}}}
\expandafter\def\csname PY@tok@s2\endcsname{\def\PY@tc##1{\textcolor[rgb]{0.73,0.13,0.13}{##1}}}
\expandafter\def\csname PY@tok@sh\endcsname{\def\PY@tc##1{\textcolor[rgb]{0.73,0.13,0.13}{##1}}}
\expandafter\def\csname PY@tok@s1\endcsname{\def\PY@tc##1{\textcolor[rgb]{0.73,0.13,0.13}{##1}}}
\expandafter\def\csname PY@tok@mb\endcsname{\def\PY@tc##1{\textcolor[rgb]{0.40,0.40,0.40}{##1}}}
\expandafter\def\csname PY@tok@mf\endcsname{\def\PY@tc##1{\textcolor[rgb]{0.40,0.40,0.40}{##1}}}
\expandafter\def\csname PY@tok@mh\endcsname{\def\PY@tc##1{\textcolor[rgb]{0.40,0.40,0.40}{##1}}}
\expandafter\def\csname PY@tok@mi\endcsname{\def\PY@tc##1{\textcolor[rgb]{0.40,0.40,0.40}{##1}}}
\expandafter\def\csname PY@tok@il\endcsname{\def\PY@tc##1{\textcolor[rgb]{0.40,0.40,0.40}{##1}}}
\expandafter\def\csname PY@tok@mo\endcsname{\def\PY@tc##1{\textcolor[rgb]{0.40,0.40,0.40}{##1}}}
\expandafter\def\csname PY@tok@ch\endcsname{\let\PY@it=\textit\def\PY@tc##1{\textcolor[rgb]{0.25,0.50,0.50}{##1}}}
\expandafter\def\csname PY@tok@cm\endcsname{\let\PY@it=\textit\def\PY@tc##1{\textcolor[rgb]{0.25,0.50,0.50}{##1}}}
\expandafter\def\csname PY@tok@cpf\endcsname{\let\PY@it=\textit\def\PY@tc##1{\textcolor[rgb]{0.25,0.50,0.50}{##1}}}
\expandafter\def\csname PY@tok@c1\endcsname{\let\PY@it=\textit\def\PY@tc##1{\textcolor[rgb]{0.25,0.50,0.50}{##1}}}
\expandafter\def\csname PY@tok@cs\endcsname{\let\PY@it=\textit\def\PY@tc##1{\textcolor[rgb]{0.25,0.50,0.50}{##1}}}

\def\PYZus{\char`\_}

\def\PYZsh{\char`\#}

\def\PYZhy{\char`\-}
\def\PYZsq{\char`\'}

\makeatother

    \makeatletter
        \newbox\Wrappedcontinuationbox 
        \newbox\Wrappedvisiblespacebox 
        \newcommand*\Wrappedvisiblespace {\textcolor{red}{\textvisiblespace}} 
        \newcommand*\Wrappedcontinuationsymbol {\textcolor{red}{\llap{\tiny$\m@th\hookrightarrow$}}} 
        \newcommand*\Wrappedcontinuationindent {3ex } 
        \newcommand*\Wrappedafterbreak {\kern\Wrappedcontinuationindent\copy\Wrappedcontinuationbox} 
        \newcommand*\Wrappedbreaksatspecials {%
            \def\PYGZus{\discretionary{\char`\_}{\Wrappedafterbreak}{\char`\_}}%
            \def\PYGZob{\discretionary{}{\Wrappedafterbreak\char`\{}{\char`\{}}%
            \def\PYGZcb{\discretionary{\char`\}}{\Wrappedafterbreak}{\char`\}}}%
            \def\PYGZca{\discretionary{\char`\^}{\Wrappedafterbreak}{\char`\^}}%
            \def\PYGZam{\discretionary{\char`\&}{\Wrappedafterbreak}{\char`\&}}%
            \def\PYGZlt{\discretionary{}{\Wrappedafterbreak\char`\<}{\char`\<}}%
            \def\PYGZgt{\discretionary{\char`\>}{\Wrappedafterbreak}{\char`\>}}%
            \def\PYGZsh{\discretionary{}{\Wrappedafterbreak\char`\#}{\char`\#}}%
            \def\PYGZpc{\discretionary{}{\Wrappedafterbreak\char`\%}{\char`\%}}%
            \def\PYGZdl{\discretionary{}{\Wrappedafterbreak\char`\$}{\char`\$}}%
            \def\PYGZhy{\discretionary{\char`\-}{\Wrappedafterbreak}{\char`\-}}%
            \def\PYGZsq{\discretionary{}{\Wrappedafterbreak\textquotesingle}{\textquotesingle}}%
            \def\PYGZdq{\discretionary{}{\Wrappedafterbreak\char`\"}{\char`\"}}%
            \def\PYGZti{\discretionary{\char`\~}{\Wrappedafterbreak}{\char`\~}}%
        } 
        \newcommand*\Wrappedbreaksatpunct {%
            \lccode`\~`\.\lowercase{\def~}{\discretionary{\hbox{\char`\.}}{\Wrappedafterbreak}{\hbox{\char`\.}}}%
            \lccode`\~`\,\lowercase{\def~}{\discretionary{\hbox{\char`\,}}{\Wrappedafterbreak}{\hbox{\char`\,}}}%
            \lccode`\~`\;\lowercase{\def~}{\discretionary{\hbox{\char`\;}}{\Wrappedafterbreak}{\hbox{\char`\;}}}%
            \lccode`\~`\:\lowercase{\def~}{\discretionary{\hbox{\char`\:}}{\Wrappedafterbreak}{\hbox{\char`\:}}}%
            \lccode`\~`\?\lowercase{\def~}{\discretionary{\hbox{\char`\?}}{\Wrappedafterbreak}{\hbox{\char`\?}}}%
            \lccode`\~`\!\lowercase{\def~}{\discretionary{\hbox{\char`\!}}{\Wrappedafterbreak}{\hbox{\char`\!}}}%
            \lccode`\~`\/\lowercase{\def~}{\discretionary{\hbox{\char`\/}}{\Wrappedafterbreak}{\hbox{\char`\/}}}%
            \catcode`\.\active
            \catcode`\,\active 
            \catcode`\;\active
            \catcode`\:\active
            \catcode`\?\active
            \catcode`\!\active
            \catcode`\/\active 
            \lccode`\~`\~ 	
        }
    \makeatother

    \let\OriginalVerbatim=\Verbatim
    \makeatletter
    \renewcommand{\Verbatim}[1][1]{%
        \sbox\Wrappedcontinuationbox {\Wrappedcontinuationsymbol}%
        \sbox\Wrappedvisiblespacebox {\FV@SetupFont\Wrappedvisiblespace}%
        \def\FancyVerbFormatLine ##1{\hsize\linewidth
            \vtop{\raggedright\hyphenpenalty\z@\exhyphenpenalty\z@
                \doublehyphendemerits\z@\finalhyphendemerits\z@
                \strut ##1\strut}%
        }%
        \def\FV@Space {%
            \nobreak\hskip\z@ plus\fontdimen3\font minus\fontdimen4\font
            \discretionary{\copy\Wrappedvisiblespacebox}{\Wrappedafterbreak}
            {\kern\fontdimen2\font}%
        }%
        
        \Wrappedbreaksatspecials
        \OriginalVerbatim[#1,codes*=\Wrappedbreaksatpunct]%
    }
    \makeatother
\begin{document}

\title{Simulating thermal qubits through thermofield dynamics: an undergraduate approach using quantum computing}

\author{G. X. A. Petronilo}
\email{gustavo.petronilo@aluno.unb.br}
\affiliation{International Center of
Physics, Instituto de F\'isica, Universidade de Bras\'ilia,\\
70910-900, Bras\'ilia, DF, Brazil}

\author{M. R. Ara\'{u}jo}
\email{matheus.araujo@ufob.edu.br}
\affiliation{Grupo de Informa\c{c}\~{a}o Qu\^{a}ntica e F\'{i}sica Estat\'{i}stica, Centro de Ci\^{e}ncias Exatas e das Tecnologias, Universidade Federal do Oeste da Bahia - Campus Reitor Edgard Santos. Rua Bertioga, 892, Morada Nobre I, 47810-059 Barreiras, Bahia, Brasil.}

\author{Clebson Cruz}
\email{clebson.cruz@ufob.edu.br}
\affiliation{Grupo de Informa\c{c}\~{a}o Qu\^{a}ntica e F\'{i}sica Estat\'{i}stica, Centro de Ci\^{e}ncias Exatas e das Tecnologias, Universidade Federal do Oeste da Bahia - Campus Reitor Edgard Santos. Rua Bertioga, 892, Morada Nobre I, 47810-059 Barreiras, Bahia, Brasil.}

\begin{abstract}
{Quantum computing has attracted the attention of the scientific community in the past few decades. The development of quantum computers promises one path toward safer and faster ways to treat, extract, and transfer information. However, despite the significant advantages of quantum computing, the development of quantum devices operating at room temperature has been compromised by the thermal decoherence process. In addition, in most undergraduate and graduate quantum mechanics courses, the study of thermofield dynamics is usually neglected. 
In this scenario, this work explores a didactic approach to simulate thermal qubit systems through Thermofield Dynamics (TFD), applied in a quantum computing setup. The results show that the Bloch sphere representation for a qubit can be written in terms of the Bogoliubov transformation, which allows a practical construction for the thermal qubits in a quantum computing setup. Therefore, this work introduces thermofield dynamics through quantum computing to teachers and curious students interested in teaching and learning this important field of studying the temperature impacts on quantum protocols using the TFD technique.}
\end{abstract}



\maketitle
\section{Introduction}

In the past few years, we have seen the emergence of several technologies based on quantum properties of advanced materials \cite{wasielewski2020exploiting,dePonte:19,Santos:20c,PRL_Andolina,Baris:20,Santos:19-a, gaita2019molecular,mezenov2019metal,cruz2022quantum,auffeves2022quantum}. However, the advantages of these quantum technologies are compromised since it is impossible to decouple the system from its surroundings and the decoherence phenomena are inevitable \cite{schlosshauer2019quantum}. Quantum features such as entanglement \cite{cruz} and coherence \cite{cruz2020quantifying}, for instance, are extremely sensitive to thermal coupling, compromising their usability at room temperature and, consequently, the development of quantum technologies under room conditions \cite{cruz,schlosshauer2019quantum,cruz2022quantum}. 

{As a consequence, the study of thermal effects on the dynamics of quantum systems has attracted the attention of the scientific community \cite{xu2022landauer,PhysRevLett.123.090402,harsha2019thermofield,lee2022variational,cartas2020hyperbolic,xu2021thermofield,PhysRevX.9.021027,lopes2020finite,weber2022non}. In this scenario, the Thermofield Dynamics (TFD) appears as a useful approach since it allows the treatment of temporal and thermal contributions equally \cite{harsha2019thermofield,lee2022variational,cartas2020hyperbolic,xu2021thermofield,miceli2019thermo,sagastizabal2021variational,zhu2020generation,xu2022landauer,prudencio2014thermofield}. 
Therefore, TFD} is a natural approach to {studying quantum states at finite temperatures.} 
{In contrast to other finite-temperature quantum field theories, which employ imaginary-time based on path integrals, the TFD formalism differs} by doubling the degrees of freedom in a Hilbert space and applying a temperature-dependent Bogoliubov transformation~\cite{santana2006thermal}. The application of higher-dimensional Hilbert spaces has been used for simplifying quantum logic in the construction of  quantum circuits \cite{lanyon2009simplifying}. On the other hand, the TFD approach gives an advantage in studying gate-based quantum computers because it is a real-time operator-based approach, using Bogoliubov transformations of quantum field theory at finite temperature \cite{miceli2019thermo}. 

In this context, this theory has shown great potential for studying the thermal properties of qubit systems regarding  the development of quantum protocols at finite temperatures~\cite{xu2022landauer,miceli2019thermo,lee2022variational,sagastizabal2021variational,zhu2020generation,rowlands2018noisy,prudencio2014thermofield}. In addition, applying higher-dimensional Hilbert spaces have been used to simplify quantum logic in the construction of quantum circuits \cite{lanyon2009simplifying}.
Implementing qubits and logical gates at finite temperature using TFD {formalism has long been a question of great interest in the past few years}~\cite{xu2022landauer,lee2022variational,wu2019variational,miceli2019thermo,sagastizabal2021variational,prudencio2014thermofield,rowlands2018noisy}. As a result, the realization of simulation of thermal qubits using variational methods was developed~\cite{lee2022variational,wu2019variational,miceli2019thermo,sagastizabal2021variational}, and recently it was demonstrated that the verification of TFD doubled state on a trapped-ion quantum computer~\cite{zhu2020generation}.
Furthermore, since quantum computers can model other quantum mechanical systems, the preparation of thermal equilibrium states in quantum computers can be used to study temperature-dependent quantum phenomena \cite{xu2021thermofield,xu2022landauer}.

In this context, this paper investigates a teaching approach for simulating thermal qubit systems using Thermofield Dynamics (TFD) in a quantum computing environment. The primary objective is to present a proof-of-concept for implementing thermal qubits in a quantum computing configuration using the TFD theory, in a manner that is accessible to undergraduate students with an interest in this important and expansive field of quantum physics. We assume that the readers have knowledge of the well-known quantum mechanics and quantum computing fundamentals \cite{nielsen2002quantum}
Exploiting that the TFD formalism uses dual states to express temperature dependence \cite{miceli2019thermo}, we show how to build a Bloch sphere representation for a two-level quantum system (qubit) in terms of the Bogoliubov transformations. The thermofield-double qubits were simulated using the quantum information software development kit - \textit{IBM Qiskit},  available on the \textit{IBM Quantum Experience} platform \cite{Qiskit,jesus2021computaccao,alves2022algoritmos,oliveira2021algoritmos}, available free of charge and easily accessible to students. 

Using the TFD algebraic approach, we show the implementation of the famous quantum teleport algorithm at finite temperatures. The results show that using  Bloch sphere representation for a qubit, written in terms of the Bogoliubov transformation, allows the construction of thermal qubits in a quantum computing setup, making it more intuitive and approachable for quantum algorithm developers. Consequently, the algebraic method of TFD theory makes modeling quantum protocols at finite temperatures easier for quantum algorithm developers, lowering the number of quantum gates required. This decrease in gates simplifies quantum logic opening a large avenue for the teaching and learning of thermal effects on quantum protocols. 
Therefore, this work provides an enticing introduction to thermofield dynamics via quantum computing, aimed at undergraduate teachers, and inquisitive undergraduate and graduate students, interested in teaching and learning this significant and extensive field of researching the thermal effects on quantum protocols. The codes used in this article are accompanied by detailed protocols throughout the text\footnote{In addition, the codes used in this work are also available on GitHub \cite{git}.}, enabling readers to replicate and apply them to their own quantum computing initiatives, even if they have limited experience in scientific computing.

\section{Thermofield Dynamics}\label{TFD}

Thermofield Dynamics (TFD) is an approach to quantum field theory at finite temperature~\cite{xu2021thermofield,santana2006thermal,umezawa22,khanna2005non}. Different from other finite-temperature quantum field theories, based on fictitious imaginary times \cite{umezawa22,khanna2005non}, the TFD formalism emerges as a natural way of describing the thermal phenomena in a real-time operator-based approach through the doubling of the Hilbert space and the applying of the Bogoliubov transformations \cite{santana2006thermal,umezawa22,khanna2005non}. 

In this scenario, the tilde ($\sim$) conjugate rules give the doubling of Hilbert space, where the thermal space can be defined as ${\cal S}_T = {\cal S}\otimes\widetilde{\cal S}$, with $S$ being the standard Hilbert space and $\widetilde{S}$ being the tilde (dual) space \cite{santana2006thermal,khanna2005non}. This doubling is defined by the mapping $(\sim): {\cal S}\rightarrow\widetilde{\cal S}$, which connects each operator in $S$ to each operators in $\widetilde{S}$, such as
\begin{equation}
    {\cal A}=a\otimes 1,\qquad \widetilde{\cal A}=1\otimes a~.
\end{equation}
Thus, the relationship between the tilde $\widetilde{A}_i$ and non-tilde $A_i$ operators can be defined as:
\begin{equation}
\begin{aligned}
    ({\cal A}_i{\cal A}_j)^\thicksim &= \widetilde{{\cal A}_i}\widetilde{{\cal A}_j}, \quad (c{\cal A}_i+{\cal A}_j)^\thicksim = c^*\widetilde{{\cal A}_i}+\widetilde{{\cal A}_j}, \\ ({\cal A}_i^\dagger)^\thicksim &= \widetilde{{\cal A}_i}^\dagger, \qquad (\widetilde{{\cal A}_i})^\thicksim = -\xi {\cal A}_i,
    \end{aligned}
\end{equation}
with $\xi = -1$ for bosons and $\xi = +1$ for fermions \cite{santana2006thermal,umezawa22,khanna2005non}.

On the other hand, a Bogoliubov transformation, $U(\alpha)$, is used to introduce thermal effects by inducing a rotation between the tilde and non-tilde variables, and it is defined as
\begin{equation}
    U(\alpha)=\left( \begin{array}{cc} u(\alpha) & -v(\alpha) \\
\xi v(\alpha) & u(\alpha) \end{array} \right),
\end{equation}
where $u^2(\alpha)+\xi v^2(\alpha)=1$. The $\alpha$ parameter is defined as the compactification parameter given by $\alpha=(\alpha_0,\alpha_1,\cdots ,\alpha_{D-1})$. The temperature effect can be algebrically described by choosing $\alpha_0\equiv\beta$ and  $\alpha_1,\cdots\alpha_{D-1}=0$, where $\beta\propto\frac{1}{T}$ with $T$ being the temperature.

\subsection{Boson representation for the SU(2) algebra}\label{TFD1}

In order to introduce the concept of a thermal qubit, one can use an alternative algebraic approach to studying two-level systems as spin-$1/2$, for instance, through a boson representation of SU(2) algebra~\cite{santana2006thermal}. In this context, the ladder operators can be defined as
\begin{equation}
S^{\pm}=(\sigma^x+i\sigma^y),\qquad\text{and}\qquad S^0=\sigma^z~.\label{sp-o} 
\end{equation}
Consequently, for each spin variable, we get
\begin{eqnarray}
[S^0,S^{\pm}]=\pm S^{\pm},\qquad [S^{+},S^{-}]=2S^0~.
\end{eqnarray}
Then one can define
\begin{eqnarray}
S^{+}=a_1^\dagger a_2,\quad S^{-}=a_2^\dagger a_1,\quad \text{and} \quad S_0=\frac{1}{2}(a_1^\dagger a_1-a_2^\dagger a_2)~,
\end{eqnarray}
where $a_1$ and $a_2$ satisfying the double boson algebra
\begin{eqnarray}
[a_i,a_j^\dagger]=\delta_{i,j}
\end{eqnarray}
with $i,j=1,2$, and all other commutations being zero.

Thus, the connection to the original SU(2) algebra can be recovered assuming $n_1=s+m$, $n_2=s-m,$, where $n_i=a_i^\dagger a_i$, with $s$ and $m$ are related to the usual results
\begin{eqnarray}
\begin{aligned}
\sigma^2|s,m\rangle&=s(s+1)|s,m\rangle,\\
\sigma_z|s,m\rangle&=m|s,m\rangle
\end{aligned}
\end{eqnarray}
The rules for applying $S^{\pm}$ operators to states are as follows:
\begin{eqnarray}
\begin{aligned}
S^{-}|0,1\rangle&=0,\\
S^{-}|1,0\rangle&=|0,1\rangle,
\end{aligned}
\qquad\qquad
\begin{aligned}
S^{+}|1,0\rangle&=0,\\
S^{+}|0,1\rangle&=|1,0\rangle,
\end{aligned}
\end{eqnarray}

It is worth noting that, $S^{-}|0,1\rangle=0$, i.e. $|0,1\rangle$ is the vacuum state for $S^{-}$. Due to the fact that this represents a two-level system, one can represent a ground state and an excited state from an energetic point of view. In this context, the corresponding basis of this Hilbert space can be redefined as
\begin{equation*}
    \begin{aligned}
    |0,1\rangle=|0\rangle, \qquad
    |1,0\rangle= |1\rangle,\qquad
    \end{aligned}
\end{equation*}
where $|0\rangle$ is the ground state, and $|1\rangle$ is the excited one. Thus, one can construct the doubled Hilbert space in order to build the thermal algebra and consequently simulate thermal qubits.

\subsection{Thermo-SU(2) algebra}

In order to double the SU(2) algebra, one needs to define the set of conjugation rules, which results in the following non-vanishing commutation rules
\begin{eqnarray}
\begin{aligned}
\left[S^0,S^{\pm}\right]&=&\pm S^{\pm}\\
\left[S^{+},S^{-}\right]&=& 2S_0,\\
\left[\widetilde{S}^0,\widetilde{S}^{\pm}\right]&=&\pm \widetilde{S}^{\pm},\\
\left[\widetilde{S}^{+},\widetilde{S}^{-}\right]&=& 2\widetilde{S}_0.
\end{aligned}
\end{eqnarray}
Due to the fact that a unitary operator $U(\beta)$ can be assumed as a canonical transformation, for the algebra which characterizes the physical system, the so-called \textit{TFD thermal operators} can be defined by
\begin{eqnarray}
\begin{aligned}
S^{\pm}(\beta)&=U(\beta)S^{\pm}U^{-1}(\beta),\\
S^{0}(\beta)&=U(\beta)S^{0}U^{-1}(\beta),\\
\widetilde{S}^{\pm}(\beta)&=U(\beta)\widetilde{S}^{\pm}U^{-1}(\beta),\\
\widetilde{S}^{0}(\beta)&=U(\beta)\widetilde{S}^{0}U^{-1}(\beta),
\end{aligned}
\end{eqnarray}
where $U(\beta)$ corresponds to the \textit{Thermal Bogoliubov Transformation}.
As a consequence, the thermal operators fulfill the following requirement of destroying the so-called thermal vacuum states $|0(\beta)\rangle$:
\begin{equation}
S^-(\beta)|0(\beta)\rangle=0,\qquad\widetilde{S}^-(\beta)|0(\beta)\rangle=0,
\end{equation}
where
\begin{equation}
|0(\beta)\rangle=U(\beta)|0,\widetilde{0}\rangle=\frac{1}{Z(\beta)^{1/2}}\sum_{n=0}e^{\beta\frac{n\omega}{2}}\Big(S^+\widetilde{S}^+ - S^-\widetilde{S}^-\Big)^n|0,\widetilde{0}\rangle,\label{thermal}
\end{equation}
with $Z(\beta)=1+e^{-\beta\omega}$ being the canonical partition function, and
\begin{equation}
    U(\beta)=\frac{1}{Z(\beta)^{1/2}}\sum_{n=0}e^{-\beta\frac{n\omega}{2}}\Big(S^+\widetilde{S}^+ - S^-\widetilde{S}^-\Big)^n~.
    \label{bogo}
\end{equation}

The thermal vacuum states $|0(\beta)\rangle$ can be defined as the vacuum of the field emerging in a thermal bath, being an alternative way to represent the temperature in quantum mechanics in an algebraic perspective without the need for quantum statistics \cite{santana2006thermal}. In this scenario, the Bogoliubov transformations can be seen as a rotation in the duplicated Hilbert space which leads to the creation of thermal effects \cite{umezawa22,khanna2005non}.

\section{Thermal Qubits}

Let us consider the particular case of a two-level system (qubit). From Eq.~\eqref{thermal} one can define the thermal vacuum qubit or thermal qubit \cite{xu2022landauer} as
\begin{equation}
   |0(\beta)\rangle = \frac{1}{Z(\beta)^{1/2}}\big(|0,\widetilde{0}\rangle+e^{-\beta\frac{\omega}{2}}|1,\widetilde{1}\rangle\big).\label{tlq}
\end{equation}
It is worth noting that, since the double thermofield is always in the same state as the real fermion, the mixed states $|1,\widetilde{0}\rangle$ and $|0,\widetilde{1}\rangle$ cannot be seen \cite{santana2006thermal,umezawa22,khanna2005non}.

In this regard, rewriting Eq.~\eqref{tlq} as
\begin{equation}
|0(\beta)\rangle = a|0,\widetilde{0}\rangle+b|1,\widetilde{1}\rangle,\label{bloch}
\end{equation}
leads Eq.~\eqref{bloch} to a form that resembles the geometrical representation for a pure state of a two-level quantum system, the so-called Bloch Sphere representation of a qubit \cite{nielsen2002quantum}:
\begin{equation}
   |\psi\rangle = \cos{\tfrac{\theta}{2}}|0\rangle + e^{i\phi}\sin{\tfrac{\theta}{2}}|1\rangle,
\end{equation}
where: 
\begin{eqnarray}
\begin{aligned}
    a= \cos{\frac{\theta}{2}}=\frac{1}{\sqrt{1+e^{-\beta\omega}}}, \\
    b= \sin{\frac{\theta}{2}}=\frac{e^{-\beta\omega/2}}{\sqrt{1+e^{-\beta\omega}}}~.
\end{aligned}
\label{aeb}
\end{eqnarray}

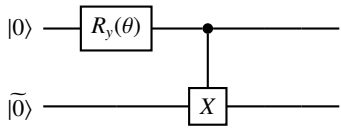
\begin{figure}[H]
    \centering
\begin{quantikz}
\lstick{$\ket{0}$} & \gate{R_y(\theta)} & \ctrl{1} & \qw & \qw &  \qw \\
\lstick{$|\widetilde{0}\rangle$} & \qw & \gate{X}& \qw & \qw & \qw 
\end{quantikz}
    \caption{Sketch of the circuit that implements the \textit{Thermal Bogoliubov Transformation} by applying the TFD approach for a single two-level quantum system. The state $\vert{0}\rangle$ is ground state in the ${\cal S}$ Hilbert space, while $|\widetilde{0}\rangle$ is the ground state of the  $\widetilde{\cal S}$ one.}
    \label{circuit}
\end{figure}

Thus, as a remarkable result, the thermal vacuum can be represented in a quantum circuit setup as sketched in   Fig.~\ref{circuit}. The circuit is created in order to implement the Bogoliubov transformation using the geometrical ideas of the Bloch sphere representation. The operator 
\begin{equation}
    R_y(\theta) = e^{-i\frac{\theta}{2} \sigma_y}=\begin{pmatrix}\cos{\frac{\theta}{2}}&-\sin{\frac{\theta}{2}}\\\sin{\frac{\theta}{2}}&\cos{\frac{\theta}{2}}\end{pmatrix}
    \label{Ry}
\end{equation}
represents a rotation of an angle $\theta$ around $Y$-direction on the non-tilde qubit $|0\rangle$, while the CNOT gate \cite{nielsen2002quantum}, controlled by the non-tilde qubit, targets the tilde one, doubling the Hilbert space in this way.

It is worth noting that this procedure is only possible due to the boson representation of SU(2) algebra reported previously in this section. This representation gives us a short circuit obtained through the analytical construction of the Bogoliubov transformation. Moreover, this result can be compared with the literature in the example reported in reference~\onlinecite{miceli2019thermo}. In this reference, the authors use a function of IBM Qiskit~\cite{Qiskit} that decomposes a given unitary matrix into quantum gates. However, it currently only works for 2-qubit, while the circuit shown in Fig.~\ref{circuit} can be implemented as a subroutine limited only by the number of qubits present in a given quantum computer. In addition, using the Bogoliubov transformation approach also differs from other results reported recently in references ~\onlinecite{sagastizabal2021variational}, and \onlinecite{wu2019variational}, since it does not use a variational approach for simulating thermal qubits. Therefore, using the algebraic approach of TFD theory can make the simulation of quantum protocols at finite temperatures more intuitive for quantum algorithm developers, reducing the number of quantum gates needed for its implementation. This reduction of the number of gates can open a large avenue for the teaching of thermal effects on quantum protocols since it reduces the cost of implementing thermal qubits, simplifying the quantum logic in the circuits \cite{wille2010reducing,langenberg2020reducing,lanyon2009simplifying}.

From Eq. \eqref{aeb}, the angle $\theta$ in the Bloch sphere representation can be defined only by the reciprocal temperature $\beta$. Thus, one can conclude that
\begin{eqnarray}
    |0(\beta\rightarrow 0)\rangle &=&\frac{1}{\sqrt{2}}\Big(|0,\widetilde{0}\rangle+|1,\widetilde{1}\rangle\Big),\label{thermo3}\\
   |0(\beta\rightarrow\infty)\rangle &=&|0,\widetilde{0}\rangle,\label{thermo1}
\end{eqnarray}
where in Eq.~\eqref{thermo3} we have the non-physical (infinity temperature) maximally thermal vacuum and Eq.~\eqref{thermo1} the non-thermal qubit. Therefore, increasing the temperature ($\beta\rightarrow 0$) will populate both ground and excited states equally, reaching the maximally thermal vacuum, Eq.~\eqref{thermo3}. Consequently, decreasing it ($\beta\rightarrow\infty$) leads the system to the non-thermal state \cite{miceli2019thermo}.

This result can be simulated by implementing the circuit sketched in  Fig.~\ref{circuit} in a quantum computing setup using the IBM Qiskit framework \cite{Qiskit}. Qiskit consists of a platform that enables high-level programming, serving as both a software development platform and a Quantum Computing language \footnote{For in-depth reading about QisKit, its initialization, commands and implementation, we recommend reading the references \cite{Qiskit,norlen2020quantum,jesus2021computaccao,alves2022algoritmos,qiskit-documentation,qiskit-textbook,qiskit-tutorials}.}. 

\subsection{Quantum algorithm}

After installing a Python language compiler and \textit{Qiskit} module\footnote{Tutorials available on the references \cite{jesus2021computaccao,qiskit-documentation,qiskit-textbook,qiskit-tutorials,github-qiskit}.}, the readers are ready to learn how to write code to simulate their thermal qubits through TFD  by constructing circuits and executing them on their own personal computers. To initiate the program, it is necessary to import the following modules into the Python environment:
\begin{itemize}
  \item[] \texttt{qiskit}: the main module, used for designing the quantum circuits and execute the quantum algorithms \cite{qiskit-textbook,qiskit-documentation,github-qiskit};
  \item[] \texttt{qiskit.circuit}: to import the sequence of coherent quantum operations used in quantum computing. \cite{qiskit-textbook,qiskit-documentation,github-qiskit};
  \item[] \texttt{numpy}: to create a mathematical workspace that involves multidimensional arrays and matrices, utilizing a vast range of mathematical functions available \cite{harris2020array};
  \item[] \texttt{matplotlib}: used for generating graphs or creating data visualizations in general \cite{hunter2007matplotlib};
  \item[] \texttt{qiskit.tools.monitor}: used to monitor the real-time execution of our algorithms in a real quantum computer by utilizing the \texttt{job{\_}monitor} function \cite{qiskit-textbook,qiskit-documentation,github-qiskit};
  \item[] \texttt{qiskit.visualization}: used to visualize the probability distributions through the function \texttt{plot{\_}histogram}.
\end{itemize}

This modules can be imported in the Python environment using the code:
\begin{tcolorbox}[breakable, size=fbox, boxrule=1pt, pad at break*=1mm,colback=cellbackground, colframe=cellborder,coltitle=black,title=Box 1: Importing the Packages]
\begin{Verbatim}[commandchars=\\\{\}]
\PY{k+kn}{from} \PY{n+nn}{qiskit} \PY{k+kn}{import} \PY{o}{*}
\PY{k+kn}{from} \PY{n+nn}{qiskit}\PY{n+nn}{.}\PY{n+nn}{circuit} \PY{k+kn}{import} \PY{o}{*}
\PY{k+kn}{import} \PY{n+nn}{numpy} \PY{k}{as} \PY{n+nn}{np}
\PY{k+kn}{import} \PY{n+nn}{matplotlib}\PY{n+nn}{.}\PY{n+nn}{pyplot} \PY{k}{as} \PY{n+nn}{plt}
\PY{k+kn}{from} \PY{n+nn}{qiskit}\PY{n+nn}{.}\PY{n+nn}{tools}\PY{n+nn}{.}\PY{n+nn}{monitor} \PY{k+kn}{import} \PY{n}{job\PYZus{}monitor}
\PY{k+kn}{from} \PY{n+nn}{qiskit}\PY{n+nn}{.}\PY{n+nn}{visualization} \PY{k+kn}{import} \PY{n}{plot\PYZus{}histogram}
\end{Verbatim}
\end{tcolorbox}

The next step is to define the rotation angle $\theta$ of the non-tilde qubit performed by the operator $R_{y}(\theta)$ in Eq. \eqref{Ry}. Thus, we can represent the thermal vacuum in quantum circuit as
 \begin{tcolorbox}[breakable, size=fbox, boxrule=1pt, pad at break*=1mm,colback=cellbackground, colframe=cellborder,coltitle=black,title=Box 2: Creating the Thermal Vacuum State]
\begin{Verbatim}[commandchars=\\\{\}]
\PY{c+c1}{\PYZsh{} Define a variable theta to be a parameter with name \PYZsq{}theta\PYZsq{}}
\PY{n}{theta} \PY{o}{=} \PY{n}{Parameter}\PY{p}{(}\PY{l+s+s1}{\PYZsq{}}\PY{l+s+s1}{theta}\PY{l+s+s1}{\PYZsq{}}\PY{p}{)}
\PY{c+c1}{\PYZsh{} Set number of qubits and classical bits to 2}
\PY{n}{qubits\PYZus{}count} \PY{o}{=} \PY{l+m+mi}{2}
\PY{n}{bits\PYZus{}count} \PY{o}{=} \PY{l+m+mi}{2}
\PY{c+c1}{\PYZsh{} Initialize a quantum circuit with two qubits}
\PY{n}{qc} \PY{o}{=} \PY{n}{QuantumCircuit}\PY{p}{(}\PY{n}{qubits\PYZus{}count}\PY{p}{,}\PY{n}{bits\PYZus{}count}\PY{p}{)}
\PY{c+c1}{\PYZsh{} Add a parametrized RX rotation on the qubit}
\PY{n}{qc}\PY{o}{.}\PY{n}{ry}\PY{p}{(}\PY{n}{theta}\PY{p}{,}\PY{l+m+mi}{0}\PY{p}{)}
\PY{n}{qc}\PY{o}{.}\PY{n}{cx}\PY{p}{(}\PY{l+m+mi}{0}\PY{p}{,}\PY{l+m+mi}{1}\PY{p}{)}
\PY{n}{qc}\PY{o}{.}\PY{n}{draw}\PY{p}{(}\PY{n}{output}\PY{o}{=}\PY{l+s+s1}{\PYZsq{}}\PY{l+s+s1}{mpl}\PY{l+s+s1}{\PYZsq{}}\PY{p}{)}
\end{Verbatim}
\end{tcolorbox}

The thermal vacuum state can be checked by measuring the corresponding qubits, the reader can measure the associated qubits and interpret the results using a classical computer. The outcomes of each qubit's measurement can be stored as classical bits (either 0 or 1) in the circuit's defined classical bits \cite{jesus2021computaccao}. Qiskit offers a powerful solution for running quantum simulations on quantum processors using the cloud access to the IBM Quantum Experience platform \cite{ibm},  through the IBM Q Provider element \cite{Qiskit,qiskit-tutorials,qiskit-textbook,qiskit-documentation}. By specifying the number of circuit repetitions (or \textit{shots}), the simulator obtain the counts for every measurement outcome in the final state. In order to utilize the IBM Q Experience through Qiskit and run projects on real quantum processors\footnote{When implementing quantum logic gates, there can be systematic errors due to various factors such as noises and other processes. These inaccuracies in qubit control are not covered in this work. For more information on how these factors affect qubit control and quantum logic gates, refer to the Qiskit \textit{User Guide} \cite{qiskit-documentation}.}, it is necessary to create a free account. This can be done by accessing the "My Account" settings in IBM Q Experience \cite{ibm}. Once the API token is obtained, IBM Q devices can be accessed from a home computer using Qiskit. To simulate the quantum circuit in the IBM quantum processor using a home-classical computer, use the following commands:
\begin{tcolorbox}[breakable, size=fbox, boxrule=1pt, pad at break*=1mm,colback=cellbackground, colframe=cellborder,coltitle=black,title=Box 3: Simulating the thermal vacuum state on a real quantum processor]
\begin{Verbatim}[commandchars=\\\{\}]
\PY{c+c1}{\PYZsh{}Adding the measurements in the circuit}
\PY{n}{qc}\PY{o}{.}\PY{n}{measure}\PY{p}{(}\PY{p}{[}\PY{l+m+mi}{0}\PY{p}{,}\PY{l+m+mi}{1}\PY{p}{]}\PY{p}{,}\PY{p}{[}\PY{l+m+mi}{0}\PY{p}{,}\PY{l+m+mi}{1}\PY{p}{]}\PY{p}{)}


\PY{c+c1}{\PYZsh{}Saving your IBM account in your device}
\PY{n}{IBMQ}\PY{o}{.}\PY{n}{save\PYZus{}account}\PY{p}{(}\PY{l+s+s1}'\PY{l+s+s1}{Users\PYZus{}Token}\PY{l+s+s1}'\PY{p}{)}

\PY{c+c1}{\PYZsh{}Executing the circuit in a real quantum computer ibmq\PYZus{}lima}
\PY{n}{IBMQ}\PY{o}{.}\PY{n}{load\PYZus{}account}\PY{p}{(}\PY{p}{)} \PY{c+c1}{\PYZsh{}Loading your IBM Account}
\PY{n}{provider} \PY{o}{=} \PY{n}{IBMQ}\PY{o}{.}\PY{n}{get\PYZus{}provider}\PY{p}{(}\PY{n}{hub} \PY{o}{=}\PY{l+s+s1}{\PYZsq{}}\PY{l+s+s1}{ibm\PYZhy{}q}\PY{l+s+s1}{\PYZsq{}}\PY{p}{)} 
\PY{n}{qcomp} \PY{o}{=} \PY{n}{provider}\PY{o}{.}\PY{n}{get\PYZus{}backend}\PY{p}{(}\PY{l+s+s1}{\PYZsq{}}\PY{l+s+s1}{ibmq\PYZus{}lima}\PY{l+s+s1}{\PYZsq{}}\PY{p}{)} \PY{c+c1}{\PYZsh{}Selecting your quantum hardware}
\PY{n}{job} \PY{o}{=} \PY{n}{execute}\PY{p}{(}\PY{n}{qc}\PY{p}{,} \PY{n}{backend}\PY{o}{=}\PY{n}{qcomp}\PY{p}{,} \PY{n}{shots}\PY{o}{=}\PY{l+m+mi}{20000}\PY{p}{)}
\PY{n}{result} \PY{o}{=} \PY{n}{job}\PY{o}{.}\PY{n}{result}\PY{p}{(}\PY{p}{)}
\PY{n}{counts} \PY{o}{=} \PY{n}{result}\PY{o}{.}\PY{n}{get\PYZus{}counts}\PY{p}{(}\PY{n}{qc}\PY{p}{)}
\end{Verbatim}
\end{tcolorbox}

\begin{figure}[]
    \centering
   \subfigure[]{\includegraphics[width=8cm]{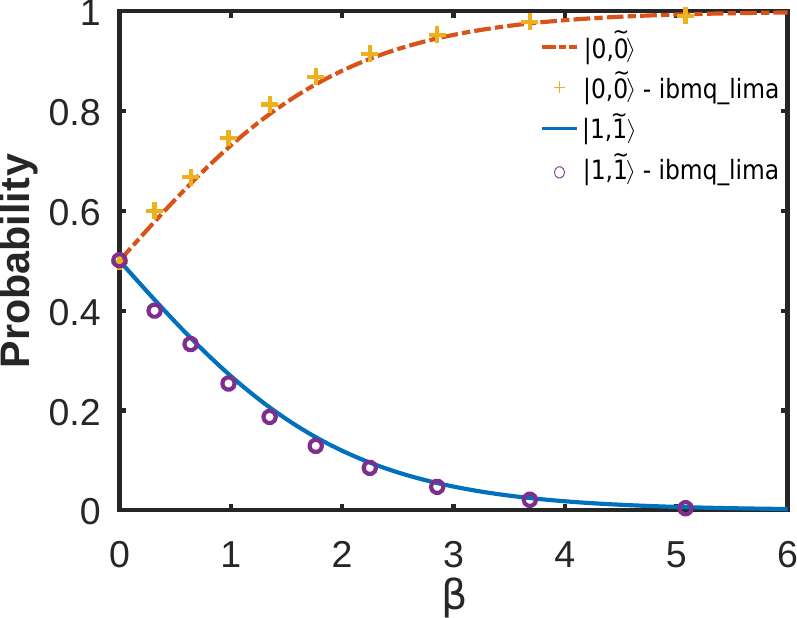}}\\
    \subfigure[]{\includegraphics[width=8cm]{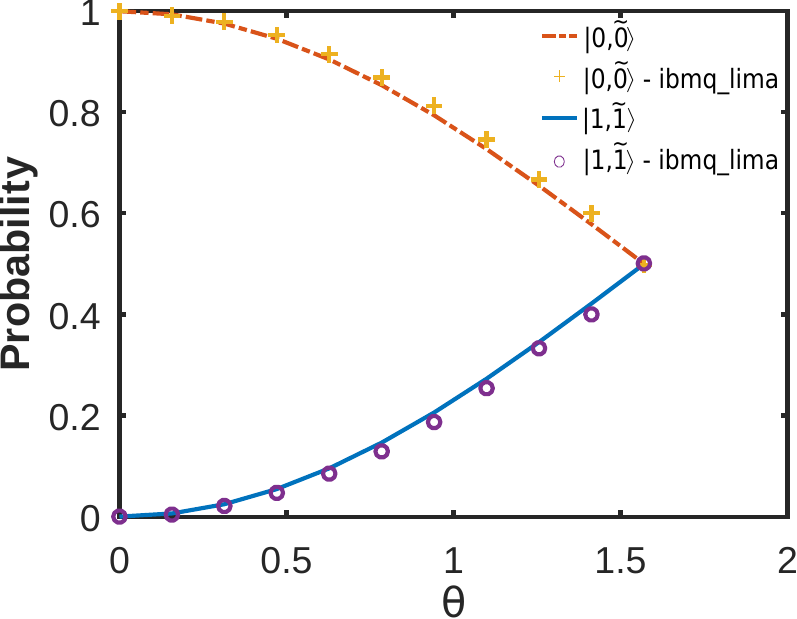}}
    \caption{(a) Probabilities of the thermal vacuum state in terms of the reciprocal temperature $\beta$ (dimensionless units). (b) Probability amplitude of the thermal vacuum in terms of $\theta$ (rad).  Dashed (red) and solid (blue) lines represent the theoretical probabilities for the ground $|0,\widetilde{0}\rangle$ and excited $|1,\widetilde{1}\rangle$, respectively. They are obtained from the square modulus of the probability amplitudes $a$ and $b$ in Eq. \eqref{aeb}.}
    \label{fig:graph1}
\end{figure}

Therefore, the thermal vacuum probabilities can be calculated by dividing the state counts by the number of shots. Fig. \ref{fig:graph1} shows the thermal vacuum probabilities as a function of the reciprocal temperature ($\beta$) and the angle $\theta$. Dashed (red) and solid (blue) lines represent the theoretical probabilities for the ground $|0,\widetilde{0}\rangle$ and excited $|1,\widetilde{1}\rangle$, and are obtained from the square modulus of the probability amplitudes $a$ and $b$ in Eq. \eqref{aeb}. The crosses and open circles represent the results obtained from the simulation of the thermal qubit in a real quantum processor, available in the backend \textit{ibmq\_lima}.

As can be seen in Fig.~\ref{fig:graph1} (a), according to Eq.\eqref{thermo1}, thermal qubit at high temperature (low $\beta$) is a superposition of the states $|0,\widetilde{0}\rangle$ and $|1,\widetilde{1}\rangle$. This superposition decays exponentially until the qubit becomes the non-thermal state $|0,\widetilde{0}\rangle$ when the temperature drops ($\beta $ rises), Eq.~\eqref{thermo3}.
In Fig.~\ref{fig:graph1} (b) we have the same prediction of Fig.~\ref{fig:graph1} (a), but now in terms of $\theta$. It should be noted that $\theta\in[0,\pi/2]$ when $\beta\in[0,\infty]$, this result is in agreement with of quantum field theories at finite temperature that see $\beta$ as a compactification of a time dimension \cite{umezawa22,khanna2005non}.

Similar results were recently shown in reference \onlinecite{miceli2019thermo} in terms of $\beta$. However, the authors used a function \textit{UnitaryGate} available in IBM Qiskit, which creates a quantum gate from a numeric unitary matrix \cite{Qiskit}. On the other hand, the result showed in Fig. \ref{fig:graph1} is directly obtained from the Bloch sphere representation in terms of the Bogoliubov transformations, obtained from the circuit shown in Fig. \ref{circuit}, in which the angle of rotation in the $R_y(\theta)$ gate defines the temperature.

\section{Quantum teleportation at finite temperature}

In this regard, one can apply the formalism shown in this section in order to simulate a quantum algorithm involving thermofield states.  As an example, we will simulate the quantum teleportation of thermofield states, a problem that has been previously discussed in reference \onlinecite{prudencio2014thermofield}. In this reference, the quantum teleportation of thermal qubits will consist of the transmission of a non-thermal qubit $|0,\widetilde{0}\rangle$ into a thermal reservoir between two parties, conventionally known as Alice and Bob, spatially separated \cite{nielsen2002quantum,prudencio2014thermofield}.  Thus in the doubled Hilbert space representation, we have $|0,\widetilde{0}\rangle\rightarrow|0(\beta)\rangle$.

Initially, Alice has only a single copy of a non-thermal qubit, and she wants to send this qubit to Bob's thermal reservoir. Thus, Alice prepares the qubit whose information will be sent and uses a second (auxiliary) thermal vacuum state
which will be maximally entangled to a third thermal vacuum state (auxiliary) belonging to Bob, who will receive Alice's non-thermal qubit. In the following, Alice performs measurements on its two thermal qubits and informs her results through a classical channel to Bob. Finally, with this (classical) information, Bob properly performs a set of (quantum) operations in your thermal vacuum state to retrieve the non-thermal qubit sent by Alice in his thermal reservoir \cite{prudencio2014thermofield}. Therefore, at the end of the teleportation, the initial non-thermal qubit $|0,\widetilde{0}\rangle$ acquires the temperature information of Bob's thermal reservoir, becoming the thermal qubit $|0(\beta)\rangle$ in the doubled Hilbert space representation of the TFD approach, Eq. \eqref{tlq}.

\begin{figure*}[ht]
    \centering
    \includegraphics[width=15cm]{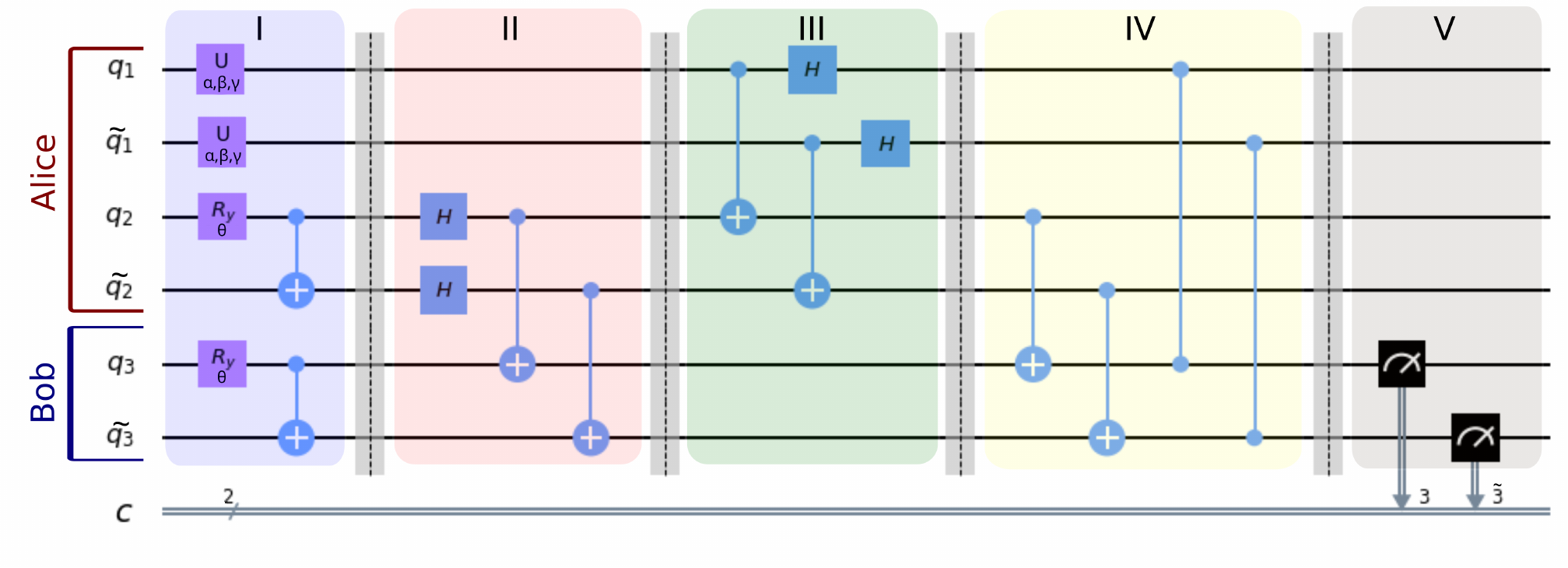}
    \caption{Teleportation circuit of a non-thermal qubit to a thermal vacuum state. The first four qubits $\{q_0,q_1,q_2,q_3\}$ belongs to Alice, while the last two $\{q_4,q_5\}$ belongs to Bob. The protocol can be divided into five main steps: (I) Preparing the thermal vacuum states; (II) Preparing the entanglement with Alice's auxiliary vacuum state $|0(\beta)\rangle_{A}$ and Bob's $|0(\beta)\rangle_{B}$; (III) Sending the non-thermal qubit; (IV) Conditioned operations targeting in Bob's state; (V) Measurement of Bob's state in order to retrieve the non-thermal qubit sent by Alice.}
    \label{teleport}
\end{figure*}

From this protocol, one can build the circuit in IBM Qiskit that simulates this procedure, as shown in Fig.~\ref{teleport}. The algorithm is divided into five main steps, described as follows. 

(I) \textbf{Preparing the thermal vacuum states}. Since the usual teleport algorithm uses three qubits \cite{nielsen2002quantum}, the TFD's version needs six qubits in order to implement Bogoliubov's transformations. The first four qubits $\{q_1,\Tilde{q_1},q_2,\Tilde{q_2}\}$ belongs to Alice, while the last two $\{q_3,\Tilde{q_3}\}$ belongs to Bob. Thus, the first step of the circuit (after calling the packages as presented in Box 1) is to create the circuit registers: 
\begin{tcolorbox}[breakable, size=fbox, boxrule=1pt, pad at break*=1mm,colback=cellbackground, colframe=cellborder,coltitle=black,title=Box 4: Creating the circuit register for the TFD teleportation algorithm]
\begin{Verbatim}[commandchars=\\\{\}]
\PY{c+c1}{\PYZsh{} Creating registers}
\PY{n}{qreg\PYZus{}q} \PY{o}{=} \PY{n}{QuantumRegister}\PY{p}{(}\PY{l+m+mi}{6}\PY{p}{,} \PY{l+s+s1}{\PYZsq{}}\PY{l+s+s1}{q}\PY{l+s+s1}{\PYZsq{}}\PY{p}{)}
\PY{n}{creg\PYZus{}c} \PY{o}{=} \PY{n}{ClassicalRegister}\PY{p}{(}\PY{l+m+mi}{2}\PY{p}{,} \PY{l+s+s1}{\PYZsq{}}\PY{l+s+s1}{c}\PY{l+s+s1}{\PYZsq{}}\PY{p}{)}
\PY{n}{circuit} \PY{o}{=} \PY{n}{QuantumCircuit}\PY{p}{(}\PY{n}{qreg\PYZus{}q}\PY{p}{,} \PY{n}{creg\PYZus{}c}\PY{p}{)}
\end{Verbatim}
\end{tcolorbox}
    
In addition, Alice's first two qubits $\{q_0,\Tilde{q_0}\}$ are prepared on a non-thermal state in the doubled Hilbert space using the general \texttt{U-gate} defined as applied in the tilde and non-tilde qubits:
\begin{eqnarray}
    \Ucal(\alpha,\beta,\gamma) = \begin{pmatrix}\cos{\frac{\alpha}{2}}&-e^{\gamma}\sin{\frac{\alpha}{2}}\\e^{\beta}\sin{\frac{\alpha}{2}}&e^{\beta + \gamma}\cos{\frac{\alpha}{2}}\end{pmatrix}
    \label{ugate}
\end{eqnarray}
Box 5 displays the code used to initialize Alice's non-thermal state.:
\begin{tcolorbox}[breakable, size=fbox, boxrule=1pt, pad at break*=1mm,colback=cellbackground, colframe=cellborder,coltitle=black,title=Box 5: Initialization of Alice's non-thermal state]
\begin{Verbatim}[commandchars=\\\{\}]
\PY{c+c1}{\PYZsh{}Alice prepares her non\PYZhy{}thermal state to be teleported}
\PY{n}{circuit}\PY{o}{.}\PY{n}{u}\PY{p}{(}\PY{n}{alpha}\PY{p}{,}\PY{l+m+mi}{beta}\PY{p}{,}\PY{n}{gamma}\PY{p}{,}\PY{n}{qreg\PYZus{}q}\PY{p}{[}\PY{l+m+mi}{0}\PY{p}{]}\PY{p}{)}
\PY{n}{circuit}\PY{o}{.}\PY{n}{u}\PY{p}{(}\PY{n}{alpha}\PY{p}{,}\PY{l+m+mi}{beta}\PY{p}{,}\PY{n}{gamma}\PY{p}{,}\PY{n}{qreg\PYZus{}q}\PY{p}{[}\PY{l+m+mi}{1}\PY{p}{]}\PY{p}{)}
\end{Verbatim}
\end{tcolorbox}
The parameters \texttt{alpha}, \texttt{beta} and \texttt{gamma} on Box 5 can be selected by the user in order to initialize the state to be teleported.

On the other hand, using the circuit shown in Fig. \ref{circuit} (code shown in Box 2), Alice prepares an auxiliary thermal vacuum state $|0(\beta)\rangle_{A}$  in qubits $\{q2, \Tilde{q_2}\}$ while Bob does the same in qubits $\{q3, \Tilde{q_3}\}$, preparing the state $|0(\beta)\rangle_{B}$. 
The following Box shows the code for this initialization of Alice's and Bob's thermal vacuum states:
\begin{tcolorbox}[breakable, size=fbox, boxrule=1pt, pad at break*=1mm,colback=cellbackground, colframe=cellborder,coltitle=black,title=Box 6: Initialization of Alice's and Bob's thermal vacuum states]
\begin{Verbatim}[commandchars=\\\{\}]
\PY{c+c1}{\PYZsh{}Alice and Bob prepares their thermal vacuum states}
\PY{n}{circuit}\PY{o}{.}\PY{n}{ry}\PY{p}{(}\PY{n}{theta}\PY{p}{,} \PY{n}{qreg\PYZus{}q}\PY{p}{[}\PY{l+m+mi}{2}\PY{p}{]}\PY{p}{)}
\PY{n}{circuit}\PY{o}{.}\PY{n}{cx}\PY{p}{(}\PY{n}{qreg\PYZus{}q}\PY{p}{[}\PY{l+m+mi}{2}\PY{p}{]}\PY{p}{,} \PY{n}{qreg\PYZus{}q}\PY{p}{[}\PY{l+m+mi}{3}\PY{p}{]}\PY{p}{)}
\PY{n}{circuit}\PY{o}{.}\PY{n}{ry}\PY{p}{(}\PY{n}{theta}\PY{p}{,} \PY{n}{qreg\PYZus{}q}\PY{p}{[}\PY{l+m+mi}{4}\PY{p}{]}\PY{p}{)}
\PY{n}{circuit}\PY{o}{.}\PY{n}{cx}\PY{p}{(}\PY{n}{qreg\PYZus{}q}\PY{p}{[}\PY{l+m+mi}{4}\PY{p}{]}\PY{p}{,} \PY{n}{qreg\PYZus{}q}\PY{p}{[}\PY{l+m+mi}{5}\PY{p}{]}\PY{p}{)}
\PY{n}{circuit}\PY{o}{.}\PY{n}{barrier}\PY{p}{(}\PY{p}{)}
\end{Verbatim}
\end{tcolorbox}
The user can define the temperature of the reservoir by selecting the value of parameter \texttt{theta} using Eq. \eqref{aeb}.

(II) \textbf{Preparing the entanglement with the auxiliary thermal vacuum states}. The next step is to entangle Alice's auxiliary vacuum state $|0(\beta)\rangle_{A}$ with Bob's $|0(\beta)\rangle_{B}$, in the doubled Hilbert Space. Box 7 shows the code that for the entanglement of the thermal vacuum states $|0(\beta)\rangle_{A}$ and  $|0(\beta)\rangle_{B}$:
\begin{tcolorbox}[breakable, size=fbox, boxrule=1pt, pad at break*=1mm,colback=cellbackground, colframe=cellborder,coltitle=black,title=Box 7: Generating entanglement between Alice and Bob's thermal vacuum states]
\begin{Verbatim}[commandchars=\\\{\}]
\PY{c+c1}{\PYZsh{}Preparing the entanglement between Alice and Bob\PYZsq{}s thermal vacuum states}
\PY{n}{circuit}\PY{o}{.}\PY{n}{h}\PY{p}{(}\PY{n}{qreg\PYZus{}q}\PY{p}{[}\PY{l+m+mi}{2}\PY{p}{]}\PY{p}{)}
\PY{n}{circuit}\PY{o}{.}\PY{n}{cx}\PY{p}{(}\PY{n}{qreg\PYZus{}q}\PY{p}{[}\PY{l+m+mi}{2}\PY{p}{]}\PY{p}{,} \PY{n}{qreg\PYZus{}q}\PY{p}{[}\PY{l+m+mi}{4}\PY{p}{]}\PY{p}{)}
\PY{n}{circuit}\PY{o}{.}\PY{n}{h}\PY{p}{(}\PY{n}{qreg\PYZus{}q}\PY{p}{[}\PY{l+m+mi}{3}\PY{p}{]}\PY{p}{)}
\PY{n}{circuit}\PY{o}{.}\PY{n}{cx}\PY{p}{(}\PY{n}{qreg\PYZus{}q}\PY{p}{[}\PY{l+m+mi}{3}\PY{p}{]}\PY{p}{,} \PY{n}{qreg\PYZus{}q}\PY{p}{[}\PY{l+m+mi}{5}\PY{p}{]}\PY{p}{)}
\PY{n}{circuit}\PY{o}{.}\PY{n}{barrier}\PY{p}{(}\PY{p}{)}
\end{Verbatim}
\end{tcolorbox}

(III) \textbf{Sending the non-thermal qubit}. Then Alice starts the process of sending the non-thermal state using his auxiliary thermal vacuum state  $|0(\beta)\rangle_{A}$, which is now entangled with Bob's $|0(\beta)\rangle_{B}$. The code for this process is displayed in the following box:
\begin{tcolorbox}[breakable, size=fbox, boxrule=1pt, pad at break*=1mm,colback=cellbackground, colframe=cellborder,coltitle=black,title=Box 8: Decoding Alice's non-thermal state into her auxiliary thermal vacuum state.]
\begin{Verbatim}[commandchars=\\\{\}]
\PY{c+c1}{\PYZsh{}Alice decodes her non\PYZhy{}termal state into his auxiliary thermal vacuum state}
\PY{n}{circuit}\PY{o}{.}\PY{n}{cx}\PY{p}{(}\PY{n}{qreg\PYZus{}q}\PY{p}{[}\PY{l+m+mi}{0}\PY{p}{]}\PY{p}{,} \PY{n}{qreg\PYZus{}q}\PY{p}{[}\PY{l+m+mi}{2}\PY{p}{]}\PY{p}{)}
\PY{n}{circuit}\PY{o}{.}\PY{n}{h}\PY{p}{(}\PY{n}{qreg\PYZus{}q}\PY{p}{[}\PY{l+m+mi}{0}\PY{p}{]}\PY{p}{)}
\PY{n}{circuit}\PY{o}{.}\PY{n}{cx}\PY{p}{(}\PY{n}{qreg\PYZus{}q}\PY{p}{[}\PY{l+m+mi}{1}\PY{p}{]}\PY{p}{,} \PY{n}{qreg\PYZus{}q}\PY{p}{[}\PY{l+m+mi}{3}\PY{p}{]}\PY{p}{)}
\PY{n}{circuit}\PY{o}{.}\PY{n}{h}\PY{p}{(}\PY{n}{qreg\PYZus{}q}\PY{p}{[}\PY{l+m+mi}{1}\PY{p}{]}\PY{p}{)}
\PY{n}{circuit}\PY{o}{.}\PY{n}{barrier}\PY{p}{(}\PY{p}{)}
\end{Verbatim}
\end{tcolorbox}

(IV) \textbf{Conditioned operations}. The following step would be for Alice to perform measurements on her qubits. Depending on the results, contact Bob through a classic channel to inform the corrections that Bob must apply in his state in order to finish the teleportation process, and he can retrieve the non-thermal qubit sent by Alice. This step can be performed through an operation conditioned to the result of Alice's measurements. However, IBM QE does not allow the implementation of this type of port conditioned to a classic channel \cite{Qiskit}. In this case, we can replace it with the \texttt{CNOT} and \texttt{Z-Controlled} gates. In this regard, Alice performs controlled gates on Bob's thermal vacuum state using her qubits as controls. Thus, we were able to modify the original circuit without changing its main objective. The following Box displays the code for this process:
\begin{tcolorbox}[breakable, size=fbox, boxrule=1pt, pad at break*=1mm,colback=cellbackground, colframe=cellborder,coltitle=black,title=Box 9: Executing Alice's conditioned gates on Bob's thermal vacuum state.]
\begin{Verbatim}[commandchars=\\\{\}]
\PY{c+c1}{\PYZsh{}Alice performs conditioned gates targeted to Bob\PYZsq{}s thermal vacuum state}
\PY{n}{circuit}\PY{o}{.}\PY{n}{cx}\PY{p}{(}\PY{n}{qreg\PYZus{}q}\PY{p}{[}\PY{l+m+mi}{2}\PY{p}{]}\PY{p}{,} \PY{n}{qreg\PYZus{}q}\PY{p}{[}\PY{l+m+mi}{4}\PY{p}{]}\PY{p}{)}
\PY{n}{circuit}\PY{o}{.}\PY{n}{cz}\PY{p}{(}\PY{n}{qreg\PYZus{}q}\PY{p}{[}\PY{l+m+mi}{0}\PY{p}{]}\PY{p}{,} \PY{n}{qreg\PYZus{}q}\PY{p}{[}\PY{l+m+mi}{4}\PY{p}{]}\PY{p}{)}
\PY{n}{circuit}\PY{o}{.}\PY{n}{cx}\PY{p}{(}\PY{n}{qreg\PYZus{}q}\PY{p}{[}\PY{l+m+mi}{3}\PY{p}{]}\PY{p}{,} \PY{n}{qreg\PYZus{}q}\PY{p}{[}\PY{l+m+mi}{5}\PY{p}{]}\PY{p}{)}
\PY{n}{circuit}\PY{o}{.}\PY{n}{cz}\PY{p}{(}\PY{n}{qreg\PYZus{}q}\PY{p}{[}\PY{l+m+mi}{1}\PY{p}{]}\PY{p}{,} \PY{n}{qreg\PYZus{}q}\PY{p}{[}\PY{l+m+mi}{5}\PY{p}{]}\PY{p}{)}
\PY{n}{circuit}\PY{o}{.}\PY{n}{barrier}\PY{p}{(}\PY{p}{)}
\end{Verbatim}
\end{tcolorbox}

(V) \textbf{Measurement in Bob's vacuum state}. Finally, Bob measures his thermal vacuum state, retrieving the non-thermal qubit sent by Alice into his thermal reservoir as the thermal qubit $|0(\beta)\rangle$, Eq. \eqref{tlq}. The following Box exhibits the code for this step:
\begin{tcolorbox}[breakable, size=fbox, boxrule=1pt, pad at break*=1mm,colback=cellbackground, colframe=cellborder,coltitle=black,title=Box 10: Measuring Bob's thermal vacuum state.]
\begin{Verbatim}[commandchars=\\\{\}]
\PY{c+c1}{\PYZsh{} Bob\PYZsq{}s measurements}
\PY{n}{circuit}\PY{o}{.}\PY{n}{measure}\PY{p}{(}\PY{n}{qreg\PYZus{}q}\PY{p}{[}\PY{l+m+mi}{4}\PY{p}{]}\PY{p}{,} \PY{n}{creg\PYZus{}c}\PY{p}{[}\PY{l+m+mi}{0}\PY{p}{]}\PY{p}{)}
\PY{n}{circuit}\PY{o}{.}\PY{n}{measure}\PY{p}{(}\PY{n}{qreg\PYZus{}q}\PY{p}{[}\PY{l+m+mi}{5}\PY{p}{]}\PY{p}{,} \PY{n}{creg\PYZus{}c}\PY{p}{[}\PY{l+m+mi}{1}\PY{p}{]}\PY{p}{)}
\PY{n}{circuit}\PY{o}{.}\PY{n}{draw}\PY{p}{(}\PY{n}{output}\PY{o}{=}\PY{l+s+s1}{\PYZsq{}}\PY{l+s+s1}{mpl}\PY{l+s+s1}{\PYZsq{}}\PY{p}{)}
\end{Verbatim}
\end{tcolorbox}

Thus, one can simulate the teleport algorithm through TFD algebraic perspective (Fig. \ref{teleport}) using IBM Qiskit. For the sake of simplicity, we chose the angles $\{ \alpha = \pi , \beta = 0,\gamma = pi$ in Eq. \eqref{ugate}, during step I (Box 4). Thus, Alice sends the non-thermal state $\vert 1,\Tilde{1} \rangle$. Moreover, considering the microwave frequency $\hbar\omega = 1.5x10^{-22}$ J in Eq. \eqref{aeb}, one can prepare the thermal vacuum states $|0(\beta)\rangle_{A}$ and $|0(\beta)\rangle_{B}$ at the finite temperature $72$ K, by selecting the parameter $\theta = \pi/3$ in the step I (Box 5). In this regard, one can simulate quantum teleportation at a finite temperature (Fig. \ref{teleport}) in both ideal and real quantum processors \cite{jesus2021computaccao}. Qiskit provides a Quantum Assembly Language (QASM) simulator for quantum simulation on personal computers \cite{Qiskit,qiskit-tutorials,qiskit-textbook,qiskit-documentation}. This tool is an integral part of the module and is specifically designed to emulate the execution of quantum circuits on a local classical processor. This enables us to emulate an ideal quantum processor without any external environmental disturbances. On the other hand, in addition to being able to simulate our quantum circuit on a numerically emulated ideal processor on a home-classical computer, we can also run our projects on real quantum processors using the IBM Q Experience through the IBM Q Provider element (as shown in Box 3). In the following, Box 11 shows the code for running the quantum teleportation at a finite temperature (Fig. \ref{teleport}) in both ideal and real quantum processors:
\begin{tcolorbox}[breakable, size=fbox, boxrule=1pt, pad at break*=1mm,colback=cellbackground, colframe=cellborder,coltitle=black,title=Box 11: Simulating quantum teleportation under finite temperature using an ideal and a real quantum processor.]
\begin{Verbatim}[commandchars=\\\{\}]
\PY{c+c1}{\PYZsh{}Ideal quantum processor}
\PY{n}{simulator} \PY{o}{=} \PY{n}{Aer}\PY{o}{.}\PY{n}{get\PYZus{}backend}\PY{p}{(}\PY{l+s+s1}{\PYZsq{}}\PY{l+s+s1}{qasm\PYZus{}simulator}\PY{l+s+s1}{\PYZsq{}}\PY{p}{)}
\PY{n}{result} \PY{o}{=} \PY{n}{execute}\PY{p}{(}\PY{n}{circuit}\PY{p}{,} \PY{n}{backend}\PY{o}{=}\PY{n}{simulator}\PY{p}{,} \PY{n}{shots} \PY{o}{=} \PY{l+m+mi}{20000}\PY{p}{)}\PY{o}{.}\PY{n}{result}\PY{p}{(}\PY{p}{)}
\PY{n}{plot\PYZus{}histogram}\PY{p}{(}\PY{n}{result}\PY{o}{.}\PY{n}{get\PYZus{}counts}\PY{p}{(}\PY{n}{circuit}\PY{p}{)}\PY{p}{)}

\PY{c+c1}{\PYZsh{}Real quantum processor ibm\PYZus{}lagos}
\PY{n}{IBMQ}\PY{o}{.}\PY{n}{load\PYZus{}account}\PY{p}{(}\PY{p}{)} \PY{c+c1}{\PYZsh{}Loading your IBM Account}
\PY{n}{provider} \PY{o}{=} \PY{n}{IBMQ}\PY{o}{.}\PY{n}{get\PYZus{}provider}\PY{p}{(}\PY{n}{hub} \PY{o}{=}\PY{l+s+s1}{\PYZsq{}}\PY{l+s+s1}{ibm\PYZhy{}q}\PY{l+s+s1}{\PYZsq{}}\PY{p}{)} 
\PY{n}{qcomp} \PY{o}{=} \PY{n}{provider}\PY{o}{.}\PY{n}{get\PYZus{}backend}\PY{p}{(}\PY{l+s+s1}{\PYZsq{}}\PY{l+s+s1}{ibm\PYZus{}lagos}\PY{l+s+s1}{\PYZsq{}}\PY{p}{)} \PY{c+c1}{\PYZsh{}Selecting your quantum hardware}
\PY{n}{job} \PY{o}{=} \PY{n}{execute}\PY{p}{(}\PY{n}{circuit}\PY{p}{,} \PY{n}{backend}\PY{o}{=}\PY{n}{qcomp}\PY{p}{,} \PY{n}{shots}\PY{o}{=}\PY{l+m+mi}{20000}\PY{p}{)}
\PY{n}{result} \PY{o}{=} \PY{n}{job}\PY{o}{.}\PY{n}{result}\PY{p}{(}\PY{p}{)}
\PY{n}{counts} \PY{o}{=} \PY{n}{result}\PY{o}{.}\PY{n}{get\PYZus{}counts}\PY{p}{(}\PY{n}{circuit}\PY{p}{)}
\end{Verbatim}
\end{tcolorbox}

\begin{figure}[h]
    \centering
    \subfigure[]{\includegraphics[width=8cm]{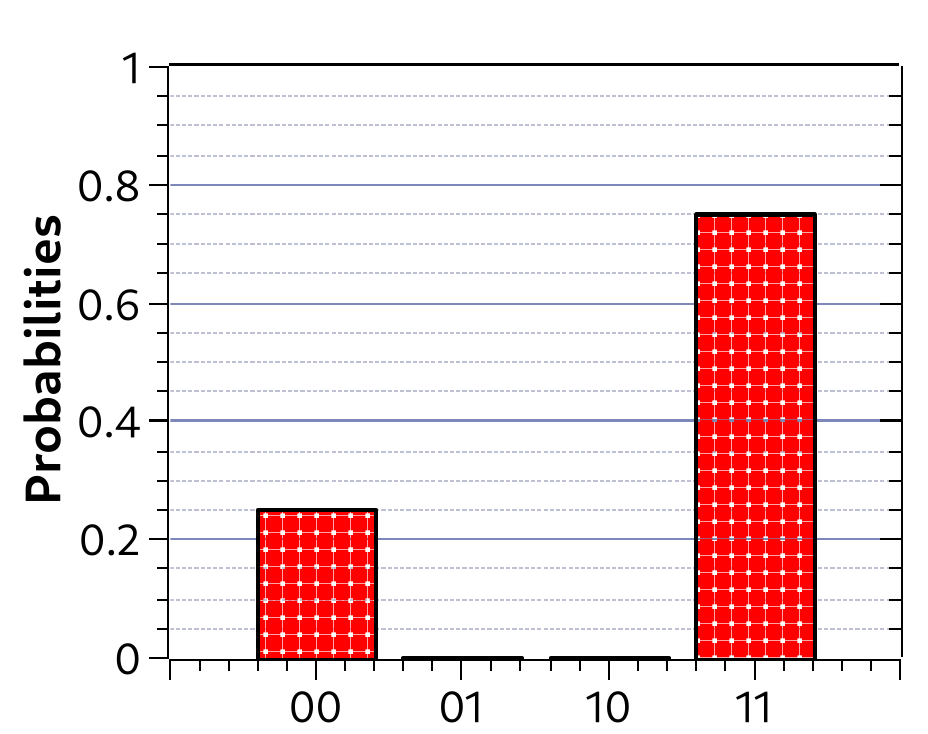}}\\
    \subfigure[]{\includegraphics[width=8cm]{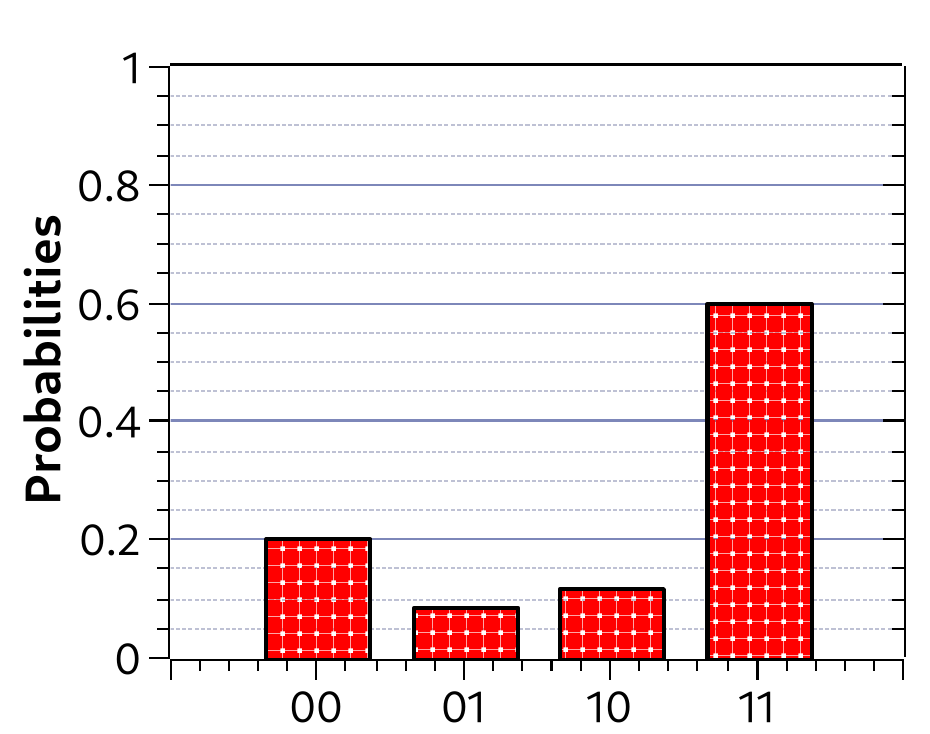}}
    \caption{Probability distribution for the teleportation algorithm at finite temperature simulated for 8000 shots in (a) an ideal quantum processor (\texttt{QASM} simulator~\cite{mckay2018qiskit}), and (b) a 7-qubit real quantum processor \textit{ibm\_oslo}. The simulation is performed considering the frequency $\omega = 1.5\cdot 10^{29}$ s$^{-1}$ and 145 K  for the temperature of the thermal vacuum states.}
    \label{teleport2}
\end{figure}

Fig. \ref{teleport2} shows the probability distributions for this teleport algorithm simulated in ideal quantum processor (\texttt{QASM} simulator~\cite{mckay2018qiskit}) and the 7-qubit real quantum processor \texttt{ibm\_lagos}\footnote{The \texttt{ibm\_lagos} processor architecture can be found on IBM QE platform \cite{ibm}. The details of the processor are beyond the scope of this paper.}. As can be seen, the state measured in Bob's vacuum state  $|0(\beta)\rangle_{B}$ was, in a good approximation, the thermal qubit state
\begin{equation}
    |1(\beta)\rangle = \sqrt{\frac{1}{4}} \vert 0,\Tilde{0}\rangle  + \sqrt{\frac{3}{4}}\vert 1,\Tilde{1} \rangle~,
    \label{tele}
\end{equation}
which is in agreement with the application of the Thermal Bogoliubov Transformation, Eq. \eqref{bogo}, on Alice's non-thermal state $\vert 1,\Tilde{1} \rangle$, with corresponding temperature $72$ K. Thus, the non-thermal state sent by Alice acquires the temperature information of the bath, becoming a thermal qubit in agreement with the presented Thermofield Dynamics model. 

The ideal processor numerically emulates the quantum circuit execution shown in Fig. \ref{teleport} (a), without any influence from external perturbations due to the inevitable coupling between quantum information processing systems and the external environment. The simulation returns the counts of each measurement in the final thermal state for the given set of 20000 shots. However, even if no decoherence act and no additional errors affect the system statistics, due to the fact that the number of shots is not large enough, one can still notice a slight statistical discrepancy between the obtained result and the theoretically expected result observed in Fig. \ref{fig:graph1} (b). On the other hand, in the real quantum processor, noise and other decohering processes lead to imprecision in the fine control of the qubits, causing systematic errors in the implementation of logic gates, which explains the appearance of the states $|0,\widetilde{1}\rangle$ and $|1,\widetilde{0}\rangle$. Nevertheless, these results show a proof of concept regarding the implementation of thermal qubits through the TFD approach. Therefore, using the Bloch sphere representation in terms of the Bogoliubov transformation allows the construction of thermal qubits in a quantum computing setup, opening a large avenue for future research toward the study of thermal effects in the development of optimal quantum protocols.

\section{Conclusion}\label{conc}

In summary, this work shows how quantum computing can be a useful tool for teaching the Thermofield Dynamics and the Blogoliobov transformations in a practical way. We show a didactic implementation of thermal vacuum qubits in a quantum computing setup, using an algebraic approach accessible to most undergraduate physics students. The Bloch sphere representation for a qubit is built in terms of the  Bogoliubov transformations, and the thermofield-double space is simulated through the IBM  Qiskit. The approach reported in this work can be implemented as a practical algebraic framework in undergraduate and graduate classrooms to study the thermal effects of the TFD approach using quantum computing. The presented examples show that constructing thermal qubits using TFD in a quantum computing setup can be more intuitive and approachable for professors, students, and even quantum algorithm developers. The presented method appears as an alternative to the complicated variational approach methods used in the literature up to date \cite{sagastizabal2021variational,wu2019variational}, reducing the number of quantum gates and processes needed for the implementation of the thermal qubits. In addition, the use of TFD theory for quantum computing opens a large avenue for physics teaching on the thermal effects on quantum protocols, which can be a path toward the understanding of the cost of implementation of thermal quantum algorithms. Therefore, this work highlights the potential of TFD algebraic perspective in quantum computing to encourage the study of optimal quantum protocols.





\section*{ACKNOWLEDGEMENTS}

{This study was financed in part by the \textit{Coordena\c{c}\~{a}o de Aperfei\c{c}oamento de Pessoal de N\'{i}vel Superior - Brasil} (CAPES) - finance code 001.}
We acknowledge the IBM Quantum services for this work.
{The views expressed are those of the authors, and do not reflect the official policy or position of IBM or the IBM Quantum team.}

\section*{COMPETING INTERESTS}

The authors declare no competing interests.

%

\end{document}